\documentclass[runningheads]{llncs}
\usepackage{graphicx}
\RequirePackage{xcolor}
\RequirePackage{ifthen,ifpdf}
\RequirePackage{colortbl}
\RequirePackage{tabularx}
\usepackage{varioref}
\usepackage{url}
\usepackage{cmap}
\usepackage[official]{eurosym}
\usepackage[T1]{fontenc}
\usepackage{times}
\usepackage{etex}
\def\href#1#2{#2}

\newcolumntype{C}{>{$}c<{$}}
\newcolumntype{P}{>{\RaggedRight}p}

\begin{document}

\title{Report on the EuDML external cooperation model}
\titlerunning{The EuDML network}
\toctitle{The EuDML external cooperation model}

\author{Thierry Bouche\inst{1} \& Ji\v r\'\i\ R\'akosn\'\i k\inst{2}}

\authorrunning{T. Bouche, J. R\'akosn\'\i k}

\institute{%
  \null\inst{1}Cellule Mathdoc (UMS 5638), Universit\'e Joseph-Fourier\\
  (Grenoble 1), B.P.\ 74, 38402 Saint-Martin d'H\`eres, France\\
  \email{thierry.bouche@ujf-grenoble.fr}\\
\inst{2}Institute of Mathematics AS CR, \v Zitn\'a 25,\\ 
  115\thinspace67 Praha 1, Czech Republic\\
\email{rakosnik@math.cas.cz}}

\maketitle

\begin{abstract}
One of the main tasks of the European Digital Mathematics Library project was to define a cooperation model with a variety of stakeholders that would allow building a~reliable and durable global reference library, aiming to be eventually exhaustive. In this paper we present the EuDML external cooperation model and the business plan as the basis for its sustainability and further development.
\end{abstract}

\section{Introduction}

The European Digital Library (EuDML)  \cite{eudml:cip,dml:Sylwestrzaketal2010,%
dml:borbinhaetal2011,eudml:project} was a project partly funded by the European 
Commission in the Competitiveness and Innovation Framework Programme, Information 
and Communication Technology Policy Support Programme, in the period from 1 February 
2010 to 31 January 2013. The EuDML project 
was explicitly envisioned as a pilot project addressing two challenges that prevented 
previous attempts towards a global digital mathematics library based on a top-down 
approach to succeed: 
\begin{enumerate}
\item Setting up the technical infrastructure to create a unified access point for 
the digital mathematical content hosted by a number of different organizations 
across various countries;

\item Defining a cooperation model with a variety of stakeholders that would allow 
building a reliable global reference library meant to run over the long term, and 
to be eventually exhaustive.
\end{enumerate}


During the three years of the project, these two goals have been pursued in parallel 
with stubbornness. On both sides the project reached clear successes and modified 
the state-of-the-art. The basic infrastructure is up and running, with a critical 
mass in content. A number of possible partners have declared interest in the 
initiative. However a lot remains to be done in order to secure these results and 
set the basis of a strong and inclusive infrastructure.

A general overview of the project and its outcome is described in \cite{TB:reviving}.  There is some slight overlap between both papers, in order to keep each of them self-contained, 
In this paper we report on the second goal mentioned above. In the first section 
we describe the situation that evolved from the project. The second section is 
devoted to the EuDML sustainability plan.

\section{EuDML at the end of project}

The EuDML has been built by the motley consortium of 16 partners from 8 European 
countries, which comprised a variety of stakeholders and expertise: universities 
(Instituto Superior T\'ecnico Lisbon, Universit\'e Joseph-Fourier Grenoble, University 
of Birmingham, Universytet Warszawski, Universidade de Santiago de Compostela, 
Ionian University Corfu, Masaryk University Brno), research institutes (Institute 
of Mathematics and Informatics BAS Sofia, Institute of Mathematics AS CR Praha, 
CNRS Grenoble), an international scientific service institution (FIZ Karlsruhe), a national research council (Consejo superior de investigaciones cient\'\i ficas Madrid), a commercial publisher (\'Edition Diffusion Presse Sciences Paris) and a private digital media agency (Made Media Ltd Birmingham), a large library (Nieders\"achsische Staats- und Universit\"atsbibliothek G\"ottingen) and, last but not least an international learned society (European Mathematical Society). The latter two had the status of associated partners to which another one came in the early stage of the project: Biblioteca Digitale Italiana di Matematica (bdim). They included digital content providers, technical developers, library, a scientific database and representatives of research communities.

EuDML as the outcome of the project aims to be a long-standing, reliable and open 
source of trusted mathematical knowledge. This implies EuDML policies, that mostly 
boil down to the following:
\begin{enumerate}
\item \textit{The texts in EuDML must have been scientifically validated and formally published.}


\item \textit{EuDML items must be open access after a finite embargo period. Once documents contributed to the library are made open access due to this policy, they cannot revert to close access later on.}


\item \textit{The digital full text of each item contributed to EuDML must be archived physically at one of the EuDML member institutions.}

\end{enumerate}

These rules ensure that the EuDML as reference library system is on a sound base, 
with ingested content available for perpetuity and openly accessible eventually. 
For this purpose the project set up a complex of frameworks, technology, workflows, 
validation procedures, schemas etc. The EuDML was built as a distributed system 
with tasks distributed among partners each of whom assumes full responsibility 
for the corresponding segment.

This worked well during the project when the partners formed a formal consortium 
tied together by a formal contract setting responsibilities with respect to the 
European Commission. The contract finished together with the project and will
be replaced with a suitable arrangement which we describe in Section 3.

The EuDML is not limited to the current digital content and the technology 
built above it. There is also important potential of further cooperation and extendibility.

On the technical front, the EuDML got expression of interest, but rather in the 
form of attracting new partners in some follow-up to the current project. On the 
political front, the EuDML got quite some awareness and support from mathematical 
societies on various occasions (The International Mathematical Union at the WDML symposium in Washington D.C., June 2012 \cite{ceicfuturedml}, the European Mathematical Society and some national mathematical societies at the EuDML workshop in Prague, 2010). On the content front, there has been a large number of discussions with potential further partners (eLibrary of the Mathematical Institute of the Serbian Academy of Sciences and Arts in Belgrade, digitized proceedings of the International Congress of Mathematics and European Congress of Mathematics, project Euclid in the Cornell University and Steklov Mathematical Institute/Russian Academy of Sciences' project Math-Net.Ru). 

An important decision that has been taken after the 6ECM round table \cite{6ecmpanel} is that the EMS Publishing House will contribute the \textit{Journal of the EMS} after a 5 
years moving wall. Work to achieve this has started, partly handled by our partner Institute of Mathematics AS CR in Prague acting here as a sponsor for EMS-ph.

A contact has been also made with JSTOR in the hope to acquire their public domain 
content and make it visible in EuDML.

The effort of the EuDML consortium does not end with creating a functional 
prototype of the Digital Mathematics Library and providing its content and services 
to the public. The true success of the project depends very much on sustainability 
and further development of the EuDML. The principal aims of sustainable EuDML services 
comprise
\begin{itemize}
\item working toward comprehensiveness, service integration, and cost efficiency of 
the EuDML services,

\item assisting in exploiting the benefits of networking for integration of digital 
library services such as sharing and enhancing data,

\item advancing cooperation between information and service providers,

\item creating and maintaining a non-profit service in the interests of the mathematics 
user community.
\end{itemize}

In order to create such sustainable service from the EuDML project, important issues 
have been assessed, namely

\begin{itemize}
\item an organizational and legal framework, which will take its roots in the EuDML 
consortium and further partners associated during its lifetime,

\item balancing costs and potential sources of revenue of running the EuDML services,

\item a common framework for dealing with intellectual properties rights and copyright issues.
\end{itemize}

\section{Principles of the EuDML Initiative}

\subsection{Organizational and legal framework}

Eleven partners of the consortium declared their will to continue in efforts to 
maintain and develop the EuDML after project's end, representing the general mathematical 
community and the core content and technology providers:

\begin{enumerate}
\item European Mathematical Society 
\item Fachinformationszentrum Karlsruhe, Zentralblatt MATH
\item Interdisciplinary Centre for Mathematical and Computational Modelling, University 
of Warsaw 
\item Cellule Mathdoc, Centre national de la recherche scientifique \& Universit\'e 
Joseph-Fourier, Grenoble 
\item University of Birmingham 
\item Institute of Mathematics and Informatics BAS, Sofia
\item Institute of Mathematics AS CR, Praha 
\item Masaryk University, Brno
\item Ionian University, Corfu
\item Societ\`a Italiana per la Matematica Applicata e Industriale, Unione Matematica 
Italiana 
\item Nieders\"achsische Staats- und Universit\"atsbibliothek G\"ottingen
\end{enumerate}

They will form an association named EuDML Initiative where the first three of them 
will assume particular r\^oles. The European Mathematical Society will provide an umbrella using its institutional authority to secure association's internal functioning and its external representation with respect to partners, other institutions and public and, in particular, to ensure that the EuDML services shall remain under control of organizations representing the public interest. For the first period of at least three years, partners no. 2 and 3 will provide human resources and machine capacities for hosting, system and service maintenance, and basic technical operations. Partner no. 4 will contribute manpower to continue enhancing the metadata quality, ingestion procedure, and will help new content providers to contribute their collections. Partner no. 7 will take care for annotations moderation and dissemination activities.

The EuDML Initiative will be established as an association without legal personality. 
The status of an association without legal personality and financial budget will 
be perfectly functional at least for the first period, during which the possibility/necessity of transforming the EuDML Initiative to another model involving legal personality and financial issues will be investigated. 

The purpose of the EuDML Initiative is to provide a Digital Mathematics Library (DML) 
for the worldwide scientific community as a public service which will help users locate the information that is distributed in various digital repositories and discover information related to their work in an efficient way, and encourage the public in using it as a public resource of knowledge which will become exhaustive and comprehensive in the field of mathematics. 

The basis for such DML is provided by the EuDML. The EuDML Initiative will 
\begin{itemize}
\item encourage content providers to join and integrate their content to the EuDML, 
\item adapt to using new information technologies and invite subjects interested in 
contributing to research and development for the continuous evolution of EuDML to 
join,
\item search for projects of research and development aiming at improving its services 
to the user community and will encourage its members to get involved in such projects. 
\end{itemize}

Membership in the EuDML Initiative will be open to any legally and contractually competent natural or legal person willing to support the objectives of the EuDML Initiative by providing

\begin{itemize}
\item digital content, i.e. integrating (at least partially) their digital collections 
of high-quality mathematical publications in the EuDML according to the EuDML guidelines 
and standards, which are based on internationally accepted standards and trends, 
adapted to the special needs of mathematical publications, and maintaining and 
expanding access to these publications through the EuDML service,
\item technological equipment and services for maintaining the EuDML central services 
and/or developing other technical services and tools to be used in the EuDML,
\item scientific, financial, strategic or political support to the EuDML Initiative and its activities.
\end{itemize}

Members will have the duty to take part in the activities necessary for the satisfaction of the objectives of the EuDML Initiative and to comply with the decisions of the bodies of the EuDML Initiative, with the statute and the applicable legislation. Members of the EuDML Initiative are not obliged to make any financial contributions. Each member cares for his own costs.

The governance and operation of the association will be organized in the following bodies: the General Assembly, the Chair of the EuDML Initiative, the Executive Board, the Scientific Advisory Board and the Technical Committee; in conducting legal or other affairs the Initiative shall be represented by the Chair of the Initiative. 

The General Assembly as the supreme decision-making body will be formed by one representative of each member. The tasks assigned to the General Assembly include decision on applications for membership, decision on expulsion of a member, election and dismissal of the Chair of the EuDML Initiative, the members of the Executive Board and of the Technical Committee from persons suggested by members of the EuDML Initiative, issuing instructions to the Executive Board, decision on modification of the statutes and the dissolution of the EuDML Initiative. 

The Executive Board will be composed of the Chair of the EuDML Initiative, the Chair of the Technical Committee and three other persons representing members of the EuDML Initiative. The European Mathematical Society's eminent r\^ole and responsibility in the EuDML Initiative will be accomplished by the fact that one member of the Executive Board will be directly nominated by the EMS Executive Committee. The Executive Board shall have general charge of all matters concerning the EuDML Initiative except for those assigned to the General Assembly, in particular, management of the current affairs including copyright and ownership management, execution of the decisions of the General Assembly, appointment of the subordinate committees entrusted with the special tasks within general framework of the association. The Executive Board will be advised by the Scientific Advisory Board and supported by the Technical Committee. 

The Scientific Advisory Board will be appointed by the European Mathematical Society 
of their representatives and other suitably qualified and recognized persons. It's 
responsibility will be ensuring the scientific quality of the DML service, and 
advising the Executive Board on scientific matters, strategic orientations and 
priorities for development of the service, taking part in the evaluation, and bringing 
in the feedback of the mathematical community. The Scientific Advisory Board works 
out recommendations for the development of EuDML with regard to the content and 
the organization of the EuDML Initiative.  

The Technical Committee will ensure the continuous technical operations of EuDML services being responsible for the technical development, technical standards and workflows, and the technical operations of the system. The Chair of the Technical Committee will be automatically member of the Executive Board, assuring cooperation of both bodies. 

\subsection{Costs and sources of revenue}

It is estimated that basic operation and maintenance of the EuDML system will require approximately 1.5 FTE and k\euro{100} yearly. This will be covered by in kind payments of the five partners indicated in Subsection 3.1. Each EuDML Initiative member will cover his local expenses which concerns, in particular, the content providers responsible for keeping and developing their repositories. It is assumed that all bodies of the association will meet via telecommunication if appropriate. Videoconferencing has proved a very efficient management tool during the EuDML project and will reduce the running costs.

The minimal level of resources corresponding to the costs indicated above will be
sufficient for the maintenance and slow development of the EuDML. However, higher revenues are desirable for a more dynamic advancement. Possible resources to cover these costs may include income from R\&D grant funding generated by the association or by its individual partners, special fees collected from the EMS member organizations, donations, financial contributions from content providers and partners of the EuDML Initiative and in kind contributions by partners.

Despite its universal usage in science, technology, education, social matters etc. mathematics has rare possibilities to generate incomes directly. All the more the EuDML as a not-for-profit establishment providing public service will have to struggle for regular resources using the potential of all partners involved and of the scientific community represented particularly by the European Mathematical Society and other mathematical societies and organizations.

\subsection{Common principles for handling data and tools}

It is very important that the potential external partners understand who the EuDML 
Initiative is and what is to be expected of a possible cooperation. Hence, the 
following principles should be adopted and published.

The content providers retain all rights pertaining to their collections. They grant 
to the EuDML Initiative the right to keep and use copies of their provided data 
for the purposes of search and retrieval display in EuDML public services. 

Each content provider may decide, whether full text will be provided to the EuDML 
Initiative and to which extent the full text might be used or distributed. Agreements 
between the EuDML Initiative and the content providers specify these and further usage 
and exploitation rights for each collection contributed to the EuDML. 

The indexing and other metadata information generated by the central services of 
EuDML (``EuDML-enhanced metadata'') is owned by the EuDML Initiative. Content providers are entitled to use and exploit copies of EuDML-enhanced metadata of those items for 
which they have provided metadata. 

Copies of metadata provided by content providers and the EuDML-enhanced metadata 
shall be kept at the sites maintained by service providers. If a service provider 
withdraws from this r\^ole, the data and respective rights and obligations stay with 
the remaining service providers. Members of the EuDML Initiative active in the research 
domain of Digital (Mathematics) Libraries are entitled to use (copies of) the EuDML-enhanced metadata for their research purposes. However, usage of a particular subset of the metadata for this purpose may be restricted by the respective information provider. Results of such research activities will be made available to the EuDML Initiative and its members. Exploitation rights for software and tools developed by the EuDML Initiative stay with the originator. However, the EuDML Initiative advocates an open-source policy for software, and encourages developers to put their developments for DML in the public domain. In case of dissolution of the EuDML Initiative, all (meta)data and related rights are to be transferred to the European Mathematical Society. 

\section{Conclusions}

The EuDML project has successfully developed a cooperation model with a variety of stakeholders has been defined for building a reliable and durable global reference 
library and a number of possible partners have declared interest in the EuDML Initiative.

The EuDML policy was developed stating three main principles: (i) the digital content 
must be scientifically validated, (ii) eventual open access, (iii) physical archiving 
of the content at one of the EuDML member institutions. 

Contacts were made with several possible external partners to ingest their digital content in EuDML. In some of these cases, the technical work has already started. 

Based on the above policies, a model of sustainable EuDML operation has been drawn 
on the basis of an association without legal personality formed by EuDML core members 
being scientifically and organizationally strong not-for-profit institutions that 
take care of the system's activity, maintenance, and of the collections both in 
terms of preservation and eventual open access provision. Three partners, EMS, 
FIZ/Zentralblatt MATH and ICM will assume particular r\^oles providing an umbrella 
securing association's internal functioning and its external representation, ensuring 
that the DML services shall remain under control of organizations representing 
the public interest, and providing human resources and machine capacities for hosting, 
system and service maintenance, and basic technical operations during the first 
mid-term period after the project end. The possibility/necessity of transforming 
the EuDML Initiative to another model involving legal personality and financial issues 
will be investigated during this period.

The EuDML Initiative will be an open, democratic association with well defined structure, distributed r\^oles and responsibilities which will allow the long-term sustainability, form the solid basis for partnership with external entities and provide condition for further development.

\bibliography{dmlbouche2}

\end{document}